\documentclass[12pt]{article}
\usepackage{graphicx}
\date{}
\begin{document}
\title{Oligo-parametric Hierarchical Structure of Complex Systems}
\author{Fariel Shafee
\\Department of Physics\\ Princeton University\\
Princeton, NJ 08540\\ USA.}

 \maketitle

\begin{abstract}
We investigate the possible origin of hierarchical structures in
complex systems describable in terms of a finite and small number of
parameters which control the behavioral pattern at each level of
organization. We argue that the limitation on the number of
important parameters at each stage is a reflection of the fact that
Thom's classification of catastrophes, i.e., qualitative changes,
involve only a few parameters. In addition, we also point out that
even in systems with a large number of components, only a few may be
of statistically great significance, just as in Zipf's law the
quantitative measure of the important collections is inversely
proportional to the rank. We then consider the concept of relative
degeneracies coming from change of resolving power, at various
scales, which too would vindicate the procedure of coarse-graining
in building up hierarchical organizations. We suggest that, similar
to the group-theoretical annihilation of dangling tensor indices due
to symmetry to minimize energy, even in more inexact contexts such
as in biology and the social sciences, similar attempts by the
system to reduce frustration may lead to cluster formation, which
are semi-closed, and let leakage interactions come into play at
larger scales.

\vspace{0.5cm}


\end{abstract}

\section{Introduction}

Virtually all complex structures have various levels of hierarchies:
cosmic clusters, planetary topography, living organisms, social
organizations, all possess this quality. In a renormalization group
\cite{RG1,RG2} approach a recurrence relation is obtained to relate
different scales and the solution of such an equation may lead to
singularities, indicating a transition point at a certain scale, but
the process of repeating such transitions at different scales
requires the formation of new sets of equations with new components,
which depend on the properties of the smaller systems.

The important point in using any equation is that it should not have
too many different terms, because otherwise it becomes incalculable,
or insoluble or nontransparent. It is debatable whether such a
criterion is an anthropic constraint created for human convenience,
but in most cases it is true that the hierarchical levels of complex
organizations are not subject to the limitations of the observing or
describing agent, but is actually observer-independent. That a large
system is not simply a collection of small subsystems has been aptly
termed as ``Many is different" by Anderson \cite{PW1}. Per Bak
\cite{PB1,PB2} created the field of self-organized criticality by
showing how complex structures can evolve from simpler units.

In this paper we shall argue that the representation  of the
behavior of each level in the hierarchy of a complex system is
related to the smallness of the number of parameters needed to
describe all kinds of basic transitions at each level, as found by
Rene Thom \cite{RT1} in his catastrophe theory. We shall consider
examples where such 'averaging' of parameters relevant to the scale
may evolve in terms of internal and external interactions. We shall
comment on the role of entropy in such a setting, where obviously a
simple additive scaling definition is irrelevant, and nonextensive
entropy of different types, such as that proposed by Tsallis
\cite{TS1,TS2,PL1}, or by us \cite{FS2,FS3,FS4} or others
\cite{KA1,KA2}  may be more appropriate.

Other treatments of the problem of the origin of hierarchy in
general, and also its metaphysical, philosophical, ecological,
social and other implications have been dealt with by several
authors over many years
\cite{SA1,WW1,PA1,AS1,BU1,DA1,HA1,HO1,SK1,NP1,NP2}

In our approach we shall refer to the semi-open nature of each level
of the hierarchy, with possible transfers of some smaller components
among the units or with the environment, while retaining the basic
qualitative integrity of the units. In the next section we review
very briefly the basic types of bifurcations which are topologically
different. In section 3 we discuss qualitatively the statistical
nature and probabilistic distribution of control parameters at any
scale and  the evolution of hierarchical structures with parameters
relevant to that scale. In section 4 we consider how even with
parameters of different magnitudes the conflicting effects of
opposing interactions can size-limit the size of clusters at any
level. In section 5 we argue why only a few effective parameters can
be expected to play the dominant role in the dynamics of a system at
a particular scale, from smoothness and symmetry considerations. In
section 6 we discuss how self-similar regimes in scale change
transformations  can produce fractal structures. In section 7 we
investigate the relevance of nonextensive entropy in describing the
pertinent information content at each scale. Lastly, in the
concluding section, we summarize our arguments and outline some
related work in progress.

\section{Oligo-parametric Control and Catastrophe Theory}

It was shown by  Thom \cite{RT1,TP1} purely from topological
analysis that all discrete changes in forms of objects can result
from only seven classes of functions, and the associated transitions
were called by him 'catastrophes'. All other functions producing
sudden changes could be associated with one of these basic functions
and can be transformed into them by differentiable mappings. Table
\ref{T1} gives the simplest forms of these functions.

\begin{table}
\begin{center}
\caption{\label{T1}Types of catastrophes and corresponding control
parameters $a,b,c,d$} \vspace{0.3cm}
\begin{tabular}{|l|l|}
\hline  \hline
equation & catastrophe  \\
\hline \hline
$x^3+ a x$ & fold\\
\hline
$x^4+ ax^2 + b x$ & cusp/Riemann-Hugoniot \\
\hline
 $x^5 + a x^3 + b x^2 + c x$ &  swallow-tail \\
 \hline
$x^3 + y^3 + a x y + b x + c y$ & hyperbolic umbilic \\

\hline $x^3/3- xy^2 + a(x^2+y^2) + bx + cy$ & elliptic umbilic \\
\hline
$x^5 + a x^4 + b x^3 + c x^2 + d x$ & butterfly \\
\hline
$x^2 y + y^4 + a x^2 + by^2 + cx + dy$ & parabolic umbilic \\
\hline
\end{tabular}
\end{center}
\end{table}

\begin{figure}[th]
\begin{center}
\includegraphics[width=8cm]
{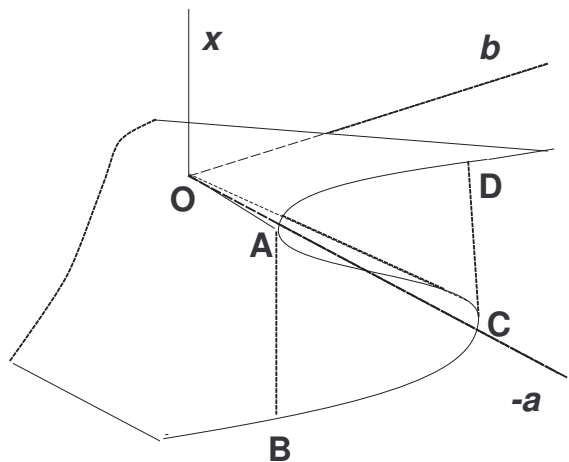} \caption{\label{Fg1}The cusp catastrophe depends on two
parameters $a$ and $b$ only and consists of two folds, where sudden
jumps take place, the middle part being a repeller. }
\end{center}
\end{figure}

The 'fold catastrophe', the simplest, is the most familiar, and is
found in many forms of phenomena in completely different contexts.
It depends on only one parameter. When the control parameter reaches
a pre-assigned value, zero in the simplest case in the equation of
the Table 1, the system bifurcates, otherwise it retains its unique
integral structure. In the cusp catastrophe, the fold is included as
a subset (Fig. \ref{Fg1}). Here one has a pair of bifurcation
points, and the system jumps from one step to the other on reaching
the critical value of a parameter $b$, provided another master
parameter $a$ is in a domain to permit the transition. Details of
applications  of catastrophic transitions may be found in many books
\cite{TP1}.

Even the most complicated catastrophe in the Table involves only
four parameters. That too involves a hierarchy of more elementary
catastrophes with some of the parameters fixed, and some varied, and
the most elementary fold is always the eventual end-product.

The complexity of higher organisms show more than a few
differentiations. But the variety of structure in such systems can
be classified at different levels with only a few types. The cell
has a cell membrane, usually also a nuclear membrane, organelles,
cytoplasmic fluid, ions,  proteins and a few other classes of
constituents. Tissues form the next level, with many cells of a
particular kind, and the kinds of such tissues is also not very
large. Then at the level of organs, individuals, species etc. we are
always dealing with oligo-differentiated systems. Hence, our
reference to small number of parameters of catastrophe theory must
be taken in the context of each level separately, though the
parameters of the successive levels may be related with one another
and with the relevant environments.

\section{Parameter Pre-Calculus}

Wilson showed in the renormalization group approach \cite{RG1} how
block spins can be created from basic spin elements with change of
scale, and how their interactions could be approximated. Given a set
of attributes with many components, as might be relevant in a more
general context than the quantum spins of condensed matter (see e.g.
\cite{FS1} for spin-glass model of social systems), the
combinatorics of formation of the ``blocks" has to be accomplished
with caution and good judgement. In physical systems, often a
group-theoretical foundation is a good guide. Singlets and lower
dimensional tensor representations are more prominent than systems
with simply additive and runaway parameters with increasing block
size. For example, in particle physics \cite{GR1}, the quarks and
anti-quarks combine to form color singlets on account of the SU(3)
symmetry group of quantum chromodynamics (consisting of $3X3$
unitary matrices transforming the three ``color" types of quarks or
anti-quarks, into one another), or, despite the symmetry breaking,
particles are usually found to be in eight-dimensional
representations of ``flavor" group SU(3)\cite{GM1},  having isotopic
or charge spin and strangeness as generators of the group. In
chemistry also atomic orbitals are so re-arranged as to produce the
most symmetric molecules allowed within the asymmetric constraints
of different atomic constituents.

From the point of view of energy, the configurations that exist
belong to the least free energies, which involves both the internal
energy as well as the entropy. We shall come back to the question of
entropy in a later section. The most symmetric states contain the
least information about the constituents. A singlet state, for
example, may be the scalar product of a vector and its adjoint, with
opposite information contents, which annihilate on combination.

Hence, a hierarchical structure involves the suppression of
information formed within semi-closed clusters. The interaction
among the clusters is due to the residual symmetry breaking part of
the interaction among the components of the constituent subsystems.
This would lead to a weaker interaction among the members of the
higher level. Quarks and anti-quarks, or quark triplets forming a
quantum-chromo-singlet form the strongest  bound systems. The
nucleus composed of protons and neutrons form the next strongest.
Atoms with electrons and nuclei are even weaker. Macroscopic forms
of matter in the form of crystals or amorphous material have weaker
binding than the binding of core electrons to their own nuclei.

Most of the properties of nuclei do not depend on the details of
quark interactions. The atom usually shows only the aggregate mass
and charge dependence of the nuclei, and so on. At each level of the
hierarchy of a complex system only a few dominant parameters derived
from the constituent level are significant enough to be in control
of the state.

For the electric field due to a complex charge distribution, one can
make a multipole expansion. The strongest energy field is the
monopole field due to a single charge, and it weakens only like
$1/r$ where $r$ is the distance from the charge to the point where
the effect is measured, and if another charge of opposite polarity
arrives nearby, which is quite likely if one is available on account
of the attraction, a dipole forms. Because the effects of the two
charges do not exactly cancel out on account of their separation,
which is small but nonzero, there is a residual field that goes like
$1/r^2$. Similarly, at the next level, one obtains the field due to
quadrupoles, which comes from the residual effect of two nearby
dipoles which come close but do not exactly cancel each other
because of the short distance between them, or because of their
nonalignment, and the energy weakens more rapidly, like $1/r^3$ in
this case. Successive terms due to higher poles, i.e. higher levels
of organization, become so weak as to be of little consequence for a
general charge distribution.

As we have mentioned in Section 2, in biological systems nuclear
membranes confine to a small region the molecules bearing the
genetic code and the auxiliary proteins during most of the cell's
life. The strongly bound double helix splits when it is required to
transmit information outside. Cells of a tissue expressing the same
genes are usually bound together within a membrane. They usually act
together in a coherent fashion, as in the heart and other organs.
The organs work in co-operation to maintain the basic functions of
an organism as a whole. At each level the interactions within the
cluster are more prominent than that with other clusters.

In a social context, the family is a strongly bound cluster, though
it usually weakens with time, with children departing to form new
clusters. There exist somewhat weaker clusters of relations,
colleagues, neighbors. At each level, details of information about
the sub-level become insignificant, and the interaction becomes
weaker than the previous level of the hierarchy.

\section{Aggregate Sizes at Each Level}

When we rank clusters according to some criterion, e.g. size (city
population), or frequency of occurrence (words in use), often there
is a simple relation between the rank and the criterion of ranking.
In the generalized form Zipf's law \cite{ZI1}has the power index
structure

\begin{equation}
C(i) = constant/R(i)^a
\end{equation}

where $C(i)$ is the number criterion for the ranking (e.g. frequency
of a word  $i$), $R(i)$ is the discrete rank, and $a$ is a number
which is independent of $i$. In Zipf's original form $a$ was shown
to be equal to unity for the case of the frequency of the most
popular English words.

Let us assume for the moment that the combinatorics of the
parameters also follows such an inverse power law. Then the
importance of the combined parameters (the criterion for the
ranking) in the clusters would decrease inversely with the rank as
we go down the ranked list. After ten such combined parameters, the
order of magnitude would then be about one order of magnitude less
than the most important one, and may be negligible for qualitative
considerations.

The origin of Zipf's law is not well understood. Power-law
probability distributions are often indicative of non-extensive
forms entropy and we shall return to this point later.

However, even with the classical Boltzmann-type  distribution we see
that the importance of the sequence of the leading combinations
becomes weaker very fast, so that it would suffice to consider only
the parametric description from only a few at any level. Let us
associate an energy $E_0$ with each member  of a sub-cluster, and
let there be $N$ such members in the  ensemble. Let there be $m$
clusters composed of different numbers of sub-clusters $m_i$ with
$i=1,2,...,n$. If the energy is simply additive, these $m$ clusters
will have the probabilistic weights of

\begin{equation}
p_i = Z e^{- \beta n_i E_0}
\end{equation}

where $Z$ is the partition function, i.e. the sum of the exponential
factors for all $i$ to give a normalized probability, and $\beta$,
the inverse of temperature, represent analogous quantities in
nonphysical systems. We see that the probabilistic weights of
subclusters comprising $n_1$ and $n_2$ components at the lower level
will be

\begin{equation}
n_1/n_2 = e^{ -\beta (n_1 - n_2)E_0}
\end{equation}

which shows an exponential decrease with $n$. If $E_0$ is negative,
i.e. a binding force, then the bigger clusters will dominate in
number, and their size would increase until a counter-force emerges
from the leakage of the semi-closed bound systems to oppose the
attractive $E_0$. In magnetism demagnetizing fields evolve to create
domain walls, and prevent all  ferromagnetic field in the sample
from aligning up in the same direction.

In the nuclear system, an inherent opposing force is the repulsive
Coulomb force between the protons which eventually limits the size
of nuclei to only about a hundred protons. In a liquid variable
clusters of molecules are size-limited by the kinetic energy.   A
joint family also develops internal strifes and is size-limited.

Hence, even in an infinite bath with an unlimited supply of
components, the aggregate size at each level is likely to be limited
at some stage. If the ensemble is large but finite, then inevitably
there must be a bounding surface for each aggregate. One then has a
competition between cohesive forces which depend on the internal
co-ordination number of the components, i.e. the effective number of
neighbors with which each component interacts, which may be
additive, and be proportional to the ``phase space" volume occupied
by the cluster, and a ``surface effect" from the components at or
near the bounding surface.

If we assume a spin-glass type interaction

\begin{equation}
H = - J^{ab}_{ij} X^i_a X^j_b  + h^a_i  X^i_a
\end {equation}

with J's representing the coupling strength between sub-clusters $a$
and $b$  and characteristics $i$ and $j$, including self-interaction
with $a=b$, and $h$ representing a coupling with the environment
(e.g. external field) or simply the mean field of the agents lying
outside the cluster, which may be formed with ``spins" in the most
immediate neighborhood and are most strongly bound.  It is known
that such a system can have frustrated ``spins" $X^a_i$, if the
signs of the components of $J$ vary \cite{FS1}. In our previous work
we did not consider the question of size-limitation, but only the
dynamical behavior of the system.

If, however, the components are given mobility, unlike the fixed
position in a spin-glass system, then the system will try to
minimize the total energy by also moving the components. This would
allow the possibility of changing neighbors to minimize frustration
and the total energy. Components which are frustrated in the midst
of other components will first move to the surface, if already in,
and will form bound systems with itinerant components which match
its characteristics better than its internal companions. One can
make a simple model of the limiting size. With $H_V$  the cohesive
binding energy per volume, and $H_S$ the surface energy with greater
affinity to the outside we get the radius of a spherical cluster

\begin{equation}
R = 3 H_S/H_V
\end{equation}

 The adhesive bondage among the peripheral components will eventually
form the seed of a new cluster. Hence, a system that has internal
interactions that allow frustrations are size-limited. In an
infinite bath with all kinds of components $X^a_i$ available, this
would lead to the formation of two super-clusters, one with all the
positive J's (``anti-ferromagnetic', with opposing nearest neighbor
spins) and the other with all the negative J's (``ferromagnetic",
all spins aligned. However, at finite temperature the disruptions by
the ambient heat bath will lead to further instability and the
super-clusters will break up into smaller clusters.

\section{Parameter Oligopoly}

\subsection{Complexity from Agent Number and Attributes}

The sub-clusters $X^a_i$ with $i = 1,2,...,n$ can, in principle each
contribute $n$ parameters for the description of the system, and if
there are $N$ such sub-clusters, the total number of possible
parameters would be $P = N n$. This may be enormously big number,
because, even if the number of characteristics per sub-cluster $n$
is small, and even if the clusters are size-limited, as outlined
above, $N$ can still be quite large. However, in many circumstances,
the reduction of the number of degrees of freedom come readily from
symmetry. In physical systems the identity of the particles makes
$X^a_i$ with  all $a$ belonging to a symmetry-related set $A$
describable by the same element $X^A_i$. In a physical context
crystal symmetry reduces the d.o.f. to only a few, despite there
being $10^{23}$ constituents, or more.

However, even with identical particles and simple and exact symmetry
groups, the interaction chain can produce a vast assortment of
complicated terms such as

\begin{equation}
H_{ijk...}  = X_i X_j X_k ....
\end{equation}

i.e. a tensor of the $N$-th rank in the group space, and each of
these terms may carry its own coping parameter in general. In
physical science many-body interactions are obtained from
perturbative expansion of two, three or four-body interactions.
Baaquie has argued recently \cite{BA1} that even in a social science
context, such as the stock market movements, quantum field-theoretic
methods may be applicable. How the presence of $C$ in the domain of
influence, or neighborhood, affects $J^{AB}$ is a potentially
insoluble problem. However, if we expand the entire interaction in
ascending order of components involved

\begin{equation}
H_{total} = J^a_{ij} X^a_i + J^{ab}_{ij} X^a_i X^b_j \\
 +J^{abc}_     {ijk} X^a_i  x^b_j  X^c_k +...
 \end{equation}

 then, like expansion in a Taylor series, we may hope that higher
 terms would not matter much. This smoothness is reasonable to
 expect, as at any level of hierarchy, a cluster can have only a
 finite number of other clusters in its domain of influence, i.e. a
 finite co-ordination number, and hence, it is not expected that
 except in pathological cases more than a few components interact at
 the same time. The cohesiveness of the system comes from  serial linkages
 which need not be as regular as a uniform  inorganic crystal, but may be a
 complex polymer chain-type structure, with a variety of
 substructures, twists and turns. But such sub-structures also
 belong to a finite number of classes, as in protein structure we
 observe alpha helices  and beta sheets in addition to linear or
 simply curved sections. In  social systems also a single person
 also acts with a finite number of members in a few groups. A demagogue may appear
 to be an exception to this rule, being able to influence a large
 number of people at the same time. But in most cases he interacts intensely
 with a coterie of like-minded political elements, and the ideas he
 broadcasts are not his alone. His companions also interact with
 similarly oriented groups, and the hierarchy exists to grass-root
 levels, and leaders at various levels of the hierarchy simply
 represent the ``block spins" at the corresponding scale.

 \subsection{Symmetry and Semi-open Systems}

 In particle physics there is a hierarchy of four interactions. The
 strongest is among quarks and anti-quarks and is short range because of
 saturation, i.e. color singlets must be produced at short ranges,
 with color field lines emerging and ending on complementary
 components, leaving no leakage for interaction with distant
 particles. Photons can leak out to infinite distances in
 electromagnetic interactions when isolated particles are
 considered, but in a plasma state, there is Debye shielding with a
 finite range of the interaction, with opposite charges almost  neutralizing
 each other within the Debye length, leaving an exponential tail to
 act at distances. The weak interactions associated usually with
 neutrinos are mediated by heavy W and Z particles and act almost
 locally. It is known from the electro-weak theory of Salam,
 Weinberg and Glashow \cite{GR1} that this interaction is only a component of a
 unified  electro-weak interaction. There is an even weaker
 interaction that violates charge conjugation and parity symmetries
 in decays of K mesons, and that too is local. The lesson to be
 learnt here is that not all local forces may be equally strong, or
 even of the same order of magnitude, if by ``locality" we restrict
 our attention only to physical co-ordinates. There is an internal
 space corresponding to symmetry groups of the particles and
 particles carrying tensor indices of different generators of the
 symmetry group and subgroups the group may interact with different
 strengths.  String theory \cite{ST1,ST2} attracted attention
 by expanding our known four-dimensional world into a
 many-dimensional world with only four retained as our familiar
 space-time and the others compactified to produce internal symmetry
 groups.

 This suggests that the dimensions may also be clustered in separate sets --
 time joining the three Galilean space dimensions in relativistic
 physics,and then the internal dimensions with associated groups with added generators -
 isospin, strangeness, charm, beauty (bottom), truth (top),
 colors etc. However the mechanism of symmetry-breaking to yield the observed
 symmetry breakings has not yet been provided by string theory, and the differentiation
 into normal space-time and more complicated internal space in terms of orbifolds
 or Calabi-Yao manifolds\cite{ST1} is done in an {\em ad hoc} manner. It is not
 clear what the basis of dimensional differentiation might be. A
 Hamiltonian in the two-dimensional space of a string, with the
 higher dimensional space-time expressed as the target-space,  fails
 to indicate the actual mechanism of the origin, if
 symmetry-breaking, which, apart from the different strengths of the
 interactions, should also give the breaking of the
 degeneracy of the normal modes of vibrations of the strings, which
 would be reflected in the great difference in masses of particles
 belonging to the same mode of vibration.

 However, assuming a separation of the internal co-ordinates from
 space-time as a phenomenological fact, we note that particles
 appear  only in the lowest irreducible representations (irreps) of
 the group. Gell-Mann \cite{GM1}, on discovery of the strangeness quantum
 number and hence, the expansion of the isospin group SU(2)  to
 $SU(3)_{flavor}$ named the principle the ``Eight-fold Way", because
 he saw the dominance of the low {\bf 8}-dimensional  irreps among the
 known  strongly interacting particles. Low dimensional
 representations carry lower number of tensorial indices from the
 constituents. A quark-antiquark  pair, with complementary indices,
  constitutes one such low dimensional representation, giving a
  meson. A three quark system gives a baryon, again belonging to two
  simplest representations, ${\bf8}$ or ${\bf10}$
  dimensional. In terms of the chromodynamic SU(3) group \cite{GR1} , both are
  singlets, and hence expected to be the most dominant clusters in
  the hierarchies as the Hamiltonian gives the lowest eigenvalues and
  most stable structures with no dangling (leaking out of the
  semi-closed system) indices, and by our ansatz of smoothness, it
  would also prefer such structures to be composed of the smallest
  number of elements. In other words , given the ensemble

  \begin{equation}
 \{q \bar {q}\} = \{q_a, q_b, ...q_z, \bar{q^a},\bar{q^b},...,
 \bar{q^z}\}
\end{equation}

where $a,b, ...$ are color indices, then they will form a structure
with pairs forming color-singlets at the nearest neighborhood, i.e
bound pairs

\begin{equation}
\left \{ (q_a \bar{q^a}), (q_b \bar{q^b}), ...,(q_z \bar{q^z})\right
\}
\end{equation}

rather than any other structure with frustrated pairs or clusters,
with uncompensated indices. However, virtual gluons would still
exchange between neighboring pairs at the next level of hierarchy,
because the pairing is not permanently saturated and stable in the
quantum sense. For very brief durations in keeping with the
uncertainty principle, a ``perfect couple" may deviate from
faithfulness and develop relations with neighbors of the opposite
color.

In atoms too, noble gases form nearly perfect closed systems with
the electron shells covering the nuclei, and thus no molecules are
formed. But here too the very close proximity of another atom
deforms the wave functions, i.e. the charge distributions, so that
the neutral atoms become dipoles and the $~ 1/r^6$ induced dipole
induced dipole interaction comes into play, and with sufficient
coercion (pressure) from an external system, these nearly closed
units can even form crystals. In biological systems the cell is a
fairly closed system, but must interact with the other cells in the
neighborhood in the same tissue. The intracellular activities are
much more dominant than intercellular ones, but here we must make a
departure from minimization of free energy as the criterion of
``dominance", and refer to survivality, which is the biological
equivalent of stability, which subcellular and inorganic
organizations strive for by minimizing free energy.

At macroscopic scales we do not usually encounter exact or even
approximate symmetry groups. However, as we have shown \cite{FS1}
the coupling of correlation matrix $J$ still exists, connecting
different agents and attributes to give a quantity (e.g.
dis-satisfaction, negative utility etc.) whose minimization can
describe the dynamics. This may also be in part a reflection of
describing agents as pure members of a symmetry group, and not
necessarily the absence of any groups. Lessons from particle physics
suggest that while enlarging a group allows putting in more
particles and interactions within the framework of the same bigger
group, such a procedure may be quite useless when the components
break the symmetry by large amounts, because then perturbative
calculations become unusable. In the social system context, there is
virtually no exact symmetry, and though the attribute labels $i$ and
the  agent labels $a$ form matrices, the difference between any two
human beings, even with identical education, cultural values,
ethnicity and other possible discriminating factors, remain far more
substantial than, say, the differences between ortho and para
hydrogen, which do not count in most contexts where hydrogen is
taken as a gas with a single component. Nevertheless, even without
closed manageable symmetry algebras, with some fuzzying of the
concepts of adjoint (complementary partners) irreps, and of the
closure of the generators of the transformations of the various
attributes, can probably provide us with methods which are
approximate but similar to those used in the stricter regime of the
physical sciences. In case of fuzzy classification, the attribute
index $i$ can represent discrete spin-type classes in place of exact
value of a continuous quantity, e.g. $X^i= 1,2,3$ may be identified
with three different broad classes of any attribute, instead
labeling by a continuous number and we can expect that associations
might form between similarly indexed agents, ie correlations
$J^{ab}_{ii}$ may be much stronger than $J^{ab}_{ij, i \neq j}$,
i.e. the correlation matrix may be nearly diagonal in attribute
space.

\begin{figure}[th]
\begin{center}
\includegraphics[width=8cm]
{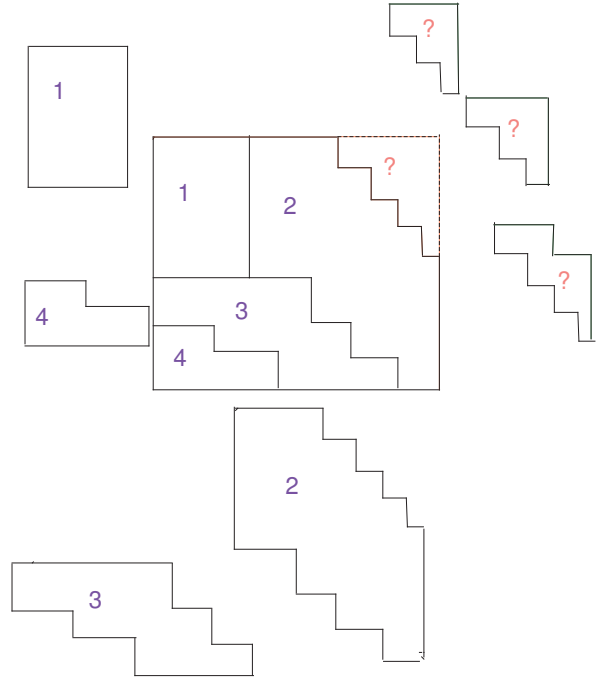} \caption{\label{Fg2}A schematic drawing showing
symmetry breaking in a cluster due to non-availability of one
appropriate component.Even with all particles identical symmetry
breaking may be inevitable as in the impossibility of arranging a
group of marbles into an exactly spherical cluster. }
\end{center}
\end{figure}

In Fig. \ref{Fg2} we indicate how a cluster can be partially
frustrated, with broken symmetry, because the right component to
complete the symmetry may not be available. In biological processes
involving enzymes meta-stable composites form with imperfect
matching, which is vital for the dynamics of life.

\section{Fractal Regimes}

As clusters collect to make bigger ones, the block spins may be
calculable in a simple fashion, and that may lead to a simple
fractal behavior \cite{MA1} of the coupling. Let us consider the
Hamiltonian at the n-th scale

\begin{equation}
H_n = J_{n,ab} s^a_n  s^b_n
\end{equation}

where we have suppressed the attribute indices $i$ for simplicity.
At the next scale level we may have a self-similar relation

\begin{equation}
H_{n+1} = J_{n+1,AB} S^A_{n+1} S^B_{n+1}
\end{equation}

However, as we have explained above, the block spins $S^A$ available
for interaction the $(n+1)$-th level may be the
frustrated/unsatiated left-overs from the semi-closed  $n$-th scale,
and hence, if there are $p$ agents in a block when going from the
$n$-th to the $(n+1)$-th level, then  $S^A$ will not be equal to $ p
s^a$, but only a fraction $p^d$ of the previous scale, with $d$ less
than unity, for example, if we have only surface elements to
consider, as argued previously, then $d=2$, with the consequence
that

\begin{equation}
J_{n+1,AB}  = p^{2d} J_{n,ab}
\end{equation}

showing a clear fractal property of the coupling with scale change.
However, other effects, such as an external field or complications
of interactions near the boundary may confine such self-similarity
only to limited domains of the scale, and there may be qualitative
transitions from one form of interaction to another, with the
interplay of even two parameters, as we have remarked in the context
of catastrophe theory.

\section{Complexity, Entropy and Parameters}

Kolmogorov measure of complexity \cite{KO1} is defined as the
minimal number of bits needed to write an algorithm that can
describe the system mathematically. Given a system with $10^{23}$
molecules in a liquid with ever changing co-ordinates, a literal
interpretation of this definition will give it such a large entropy
that it might be more like a fully random system, than even a
chaotic one, for chaos can be produced only from a simple set of
rules such as the Lorenz equations \cite{LO1}, or the logistic map
\cite{CH1,CH2}.

If we think in terms of a hierarchical structure, with the
description of different levels restricted to the finite number of
control parameters at that level, in the light of  Thom's theory,
then the Kolmogorov complexity would become a fairly manageable
number at each level.

If we try to find the Shannon entropy of a complex system, then

\begin{equation}
S = - \sum_i p_i  \log (p_i)
\end{equation}

and if there are $N$ states with almost equal probability, we get

\begin{equation}
S_{max} \sim \log (N)
\end{equation}

If there are $n$ noninteracting agents who may be distributed in
these states, we see, that $S$ scales linearly, i.e. this entropy is
extensive. If interaction is allowed, then new states may be created
and $N$ will increase.

However, if with change of scale we also filter out small
differences of the states (fine or hyperfine splittings), lumping a
number of them with similar energy into one degenerate state with a
single index $i$, then even with interaction the number of states
may not increase. The number of collective agents may be constrained
to remain the same  $n$ or of the same order of magnitude, by
choosing block spins appropriately. Hence, the entropy associated
with the larger scale would not change much from the smaller scale.
Non-extensive entropies have been proposed by various authors \cite
{TS1,TS2,KA1,KA2,FS2,FS3,FS4}, but a rigorous formulation of a
scale-independent remains an interesting problem, that may be quite
relevant in the context of our discussion about oligopoly of
parameters at all scales.

\section{Conclusions}

We have seen that hierarchical structures may from in a large
system, from interactions between the components at each stage, and
also on account of external forces. But it appears that at each
level of the hierarchy, i.e. at each scale of the system, the number
of control parameters cannot be too high. Thom's theory of
catastrophe supports this hypothesis, but such limitations also
emerge from smoothness considerations. We have argued that on
account of the semi-closed nature of the units at each level, there
is a leakage interaction among the clusters at each scale, which is
weaker than the binding energies of the sub-clusters. The
fundamental interactions of nature seem to follow such a
hierarchical pattern, as do interactions in societies.

The limitation on the number of control parameters is also justified
by approximate degeneracy of the agents at each level, as
measurements on the relevant scale would not distinguish between
fine and hyperfine splittings of states, thus curtailing the
likelihood of interactions at a smaller level augmenting
substantially the number of parameters.

Many years ago 't Hooft\cite{TH1} proposed the principle of
``naturality", where he suggested that the numbers related to
quantities we have to contend with are usually in keeping with other
numbers on which they depend. In this work we have argued that at
any scale a system can similarly be described by quantities which
are related to that scale, and hence at all scales we can expect to
get similar numbers for quantities defined for that scale, and also
the number of relevant quantities to give a broad description of the
system {\em at that scale} may be a reasonably small number, though
finer differences can become important at some level of the
hierarchy.

In a later work we shall dwell upon other interesting aspects of the
dynamics of the domain of influence, e.g. the mobility of the
(semi-)degenerate components, and the relative perceptions of time
by different subsystems evolving in interactions and in exchanges of
components between clusters. We shall also consider the question of
whether a complex abstract pattern expressible, e.g. as a
Hamiltonian with operators, may choose components from an infinite
bath to construct its physical rendition.

\section*{Acknowledgement}

I would like to thank Professor P.W. Anderson for a very useful
discussion session from which some of the ideas presented here
germinated, and also Andrew Tan of UCSF for encouragement and
feedback.


\begin{thebibliography}{99}


\bibitem{RG1} K.G. Wilson, Physica {\bf73}, 119 (1974)

\bibitem{RG2} K.G. Wilson, Nobel lecture,
nobelprize.org/nobel\_prizes/\\
physics/laureates/1982/wilson-lecture.pdf


\bibitem{PW1} P.W. Anderson, ``Many Is Different" , Science
{\bf 177}, 393  (1972)

\bibitem{PB1} P. Bak, C. Tang, and K. Wiesenfeld, ``Self-Organised Criticality: An
Explanation of 1/f Noise", {\it Phys. Rev. Lett.} {\bf 59}, 381
(1980)

\bibitem{PB2} P. Bak and K. Chen, `` Self-organized criticality", {\it  Sci.
Am.} {\bf 264}, 46  (1991)


\bibitem{RT1} Rene Thom, {\it Structural Stability and Morphogenesis: An Outline of
a General Theory of Models}, ( Addison-Wesley,Reading, MA, 1989)

\bibitem{TS1} C. Tsallis, {\it J. Stat. Phys.} {\bf 52}, 479 (1988)

\bibitem{TS2} P. Grigolini, C. Tsallis and B.J. West, {\it Chaos, Fractals and
Solitons}  {\bf13}, 367 (2001)

\bibitem{PL1} A.R. Plastino and A. Plastino, {\it J. Phys. A} {\bf
27}, 5707 (1994)

\bibitem{FS2} F. Shafee, ``A new nonextensive entropy",
nlin.AO/0406044 (2004)

\bibitem{FS3} F. Shafee, ``Generalized entropy and statistical
mechanics", cond-mat/0409000 (2004)

\bibitem{FS4} F. Shafee, `` Generalized Entropies and Quantum
Entanglement", cond-mat/0410554 (2004)

\bibitem{KA1} G. Kaniadakis, ``Nonlinear kinetics underlying generalized
Statistics", {\it Physica A} {\bf 296}, 405(2001);
cond-mat/0103467(2001)

\bibitem{KA2} G. Kaniadakis, ``Statistical mechanics in
the context of special relativity", {\it Phys. Rev. E} {\bf 66},
056125, (2002); cond-mat/0210467 (2002)

\bibitem{SA1} S. Salthe, {\it Evolving Hierarchical Systems:
their structure and representation},  (Columbia U.P., NY, 1985)

\bibitem{WW1} L. Whyte, A. G. Wilson and D. Wilson (eds.),
{\it Hierarchical structures}, (American Elsevier, NY, 1969)
hierarchical thinking.

\bibitem{PA1} H.H. Pattee(ed.),  {\it Hierarchy theory: the challenge or
complex systems}, (Braziller, NY, 1973).

\bibitem{AS1} T.F.H. Allen and T. B. Starr, {\it Hierarchy: perspectives for
ecological complexity}, (University Chicago Press, 1982)

\bibitem{BU1} J. Buchler, {\it Metaphysics of natural complexes}, (Columbia
U.  P., 1966)

\bibitem{DA1} G.J. Dalenoort, {\it The Paradigm of Self-Organization
II}, (Gordon and Breach, 1994)

\bibitem{HA1} H. Haken, {\it Synergetics}, (Springer-Verlag, 1977)

\bibitem{HO1} J. Holland, {\it Hidden Order}, (Addison Wesley, 1995)

\bibitem{SK1} S. Kauffman, {\it The Origins of Order},
 (Oxford U. P., 1993)

\bibitem{NP1} G. Nicolis and I.  Prigogine, {\it Self-Organization in Non-equilibrium
Systems}, (Wiley, 1977)

\bibitem{NP2} G. Nicolis and I. Prigogine, {\it Exploring
Complexity}, (W.H.Freeman, 1989)

\bibitem{TP1} T. Poston, and I. Stewart, {\it Catastrophe: Theory and Its
Applications}, (Dover, New York, 1998)

\bibitem{FS1} F. Shafee, ``Spin-glass-like Dynamics of Social
Networks", talk at WEHIA 2005 Conference,  physics/0506161 (2005)

\bibitem{GR1} D. Griffiths, {\it Introduction to Elementary
Particles}, (John Wiley, NY, 1987)

\bibitem{ZI1} George K. Zipf, {\it Human Behaviour and the Principle of
Least-Effort}, (Addison-Wesley, Cambridge MA, 1949)

\bibitem{BA1} B.E. Baaquie, {\it Quantum Finance}, (Cambridge University
Press, Cambridge, UK, 2005)

\bibitem{ST1} M.B. Green, J. Schwarz and E. Witten, {\it Superstring Theory},
(Cambridge University Press, Cambridge, England, 1987)

\bibitem{ST2} J. Gribbin, {\it The Search for Superstrings, Symmetry,
and the Theory of Everything}, (Little Brown, London, 1998)

\bibitem{GM1} M. Gell-Mann and Y. Ne'Eman, {\it The Eightfold Way},
(Perseus Books Group, 2000)

\bibitem{MA1} B.B. Mandelbrot, {\it The Fractal Geometry of Nature},
(W. H. Freeman,NY, 1982)

\bibitem{KO1}M. Li and P.M.B. Vitányi, {\it An Introduction to Kolmogorov
Complexity and its Applications}, (Springer-Verlag, New York, 1993)

\bibitem{LO1} E.N. Lorenz,``Deterministic nonperiodic flow", {\it J. Atmos. Sci.}
{\bf 20},  130 (1963)

\bibitem{CH1} R.M. May, ``Simple mathematical models with very complicated
dynamics",  {\it  Nature}   {\bf 261}, 459 (1976)


\bibitem{CH2} M.J. Feigenbaum, ``Quantitative universality for a class of
nonlinear transformations", {\it J. Stat. Phys.} {\bf 19}, 25 (1978)

\bibitem{TH1} G. 't Hooft, Cargese Summer Institute Lectures: 1979 ;
UTRECHT Report PRINT-80-0083 (1980)

\end{thebibliography}
\end{document}